\documentclass{article}
\usepackage{amssymb}

\input{tcilatex}
\begin{document}

\title{Finite Action Principle Revisited}
\author{John D. Barrow \\
DAMTP, Centre for Mathematical Sciences\\
University of Cambridge\\
Wilberforce Rd., Cambridge CB3 0WA\\
United Kingdom}
\maketitle

\begin{abstract}
We reconsider and extend the cosmological predictions that can be made under
the assumption that the total action of the universe is finite. When initial
and final singularities in curvature invariants are avoided, it leads to
singularities in the gravitational action of the universe. The following
properties are required of a universe with finite action: Compact spatial
sections (ie a closed universe) giving a finite total lifetime for the
universe. Compactification of flat and open universes is excluded. The
universe can contain perfect fluids with $-1<p/\rho <2$ on approach to
singularities. The universe cannot display a bounce' or indefinite cyclic
behaviour to the past or the future. Here, we establish new consequences of
imposing finite action: the universe cannot be dominated by massless scalar
fields or the kinetic energy of self-interacting scalar fields or a $p=\rho $
\ perfect fluid on approach to the initial or final singularity The
ekpyrotic scenario with an effective fluid obeying $p/\rho >2$ in a closed,
flat or open universe is excluded. Any bouncing loop quantum gravity model
with indefinite past or future evolution is ruled out. The Einstein static
and steady-state universes are ruled out along with past or future eternal
inflating universes Anisotropies of Kasner or Mixmaster type cannot dominate
the dynamics on approach to singularities. This excludes density
inhomogeneity spectra versus mass, of the form $\delta \rho /\rho \varpropto
M^{-q}$, with $q>2/3.$Higher-order lagrangian theories of gravity are
significantly constrained Quadratic lagrangians are excluded with fluids
satisfying $p/\rho >-1/3$. Lagrangians with $L_{g}=R^{1+\delta }$ have
infinite actions on approach to a singularity when $2\delta (1-3\gamma
)+2-3\gamma <0$, where $\ p=(\gamma -1)\rho .$ for the fluid. As shown by
Barrow and Tipler, the Gauss-Bonnet quadratic combination causes a
cosmological action singularity even though it does not contribute terms to
the field equations. Scalar-tensor theories like Brans-Dicke dominated by
the scalar-field on approach to singularities have action singularities.
Dark energy cannot be a simple cosmological constant, as it would create an
action singularity to the future: the universe cannot be asymptotically de
Sitter as $t\rightarrow \infty $. The dark energy must be an evolving energy
density in a closed universe that produces collapse to a future singularity
and cannot be dominated by the kinetic energy of the scalar field.
\end{abstract}

\section{ \ \ Introduction \ }

Earlier, Barrow and Tipler (BT) \cite{BT1988} explored the cosmological
consequences of requiring the gravitational and matter actions of the
universe to be finite. One feature of focussing attention upon the action is
that there have been many attempts to avoid the appearance of physical
infinities ('singularities') in cosmological models by means, for example,
of dynamical bounces at finite scale factor values or past and future
asymptotes that are non-singular. Yet, typically, these attempts to avoid
singularities give rise to infinities in the action. In this paper we
briefly summarise our original finite action conjecture of ref. \cite{BT1988}
and update its consequences and predictions in the light of subsequent
developments in observational and theoretical cosmology. There has also been
a recent rediscovery of this focus on the action, with a specific emphasis
on quadratic gravity and a computation of the action between an initial time
and the present rather than for the entire spacetime, by Lehners and Stelle 
\cite{Leh}.

In what follows we shall define the finite action proposal in section 2,
followed by a series of applications to constrain the fluid content of the
universe, the requirement of no massless scalar fields in the universe, the
ruling out of anisotropy domination at singularities and constraints on the
statistics of inhomogeneities, the exclusion of sudden finite-time
singularities, and the conclusion that the dark energy cannot be contributed
by a simple cosmological constant: it must be a variable energy source in a
closed universe that ultimately recollapses to a future singularity. In
section 3 we consider the implications of finite action for modified gravity
theories that generalise the Einstein-Hilbert lagrangian to higher order in
the curvature, the Gauss-Bonnet combination, and scalar tensor theories. Our
conclusions are listed in section 4.

\section{The Finite Action Proposal}

As discussed in BT \cite{BT1988}, there are a variety of motivations for the
fundamental importance of the action, $S$. Planck appears to have been the
first 20th century physicist to argue for the primacy of the action as the
basis for physical theories \cite{planck}. In modern physics, the action is
always the starting point because it is invariant under gauge
transformations of Yang-Mills and supersymmetric fields and its importance \
in the path-integral method of quantization \cite{ram, green}. We know that
the finiteness of the action places important constraints on some theories
in Minkowski and Euclidean spaces. For example, on solutions to the
Yang-Mills-Higgs equations and the classical solutions with finite action
lead to a semi-classical approximation to the euclidean path integral that
can be analytically continued back to Minkowski space to get the physical
path integral.

In Einstein's gravitational theory the general action of
Einstein-Hilbert-York consists of three pieces (with units such that $8\pi
G=c=1$): the gravitational, $S_{g}$, matter, $S_{m},$and boundary action
terms, respectively. Therefore, we have for the universal action:

\begin{equation}
S=\frac{1}{2}\tint\limits_{M}(R+2\Lambda )\sqrt{-g}d^{4}x+\tint%
\limits_{M}(L_{m}\sqrt{-g}d^{4}x+\tint\limits_{\partial M}(\mathrm{tr}K)(%
\sqrt{\pm h)}d^{3}x,  \label{s}
\end{equation}

where $R$ is the 4-d Ricci scalar, $\Lambda $ is the cosmological constant, $%
L_{m}$ is the matter lagrangian. The boundary of $M$ is $\partial M$ where $%
\partial M$ has induced metric $h_{\mu \nu }$ and extrinsic curvature $%
K_{ab} $. The plus(minus) sign is chosen in $(\pm h)^{1/2}$ if the boundary
is spacelike (timelike) respectively. Unless we state otherwise, we will
drop the boundary term since we want the integrations to be performed over
the whole spacetime, so there will be no boundary.

The universal action will be finite if each term in eq. (\ref{s}) is finite.
Finite action requires the Universe to have compact space sections (ie be
'closed') or else the integrals over 3-space in eq. (\ref{s}) will diverge
to infinity, so for example the steady-state universe has infinite action.
While a finite spatial volume if necessary for finite action it is not
sufficient. The Einstein static universe is closed but has infinite action
when the time integration is carried out over the time-independent
quantities $R\sqrt{-g}$ and $\Lambda \sqrt{-g}$ in eq. (\ref{s}). An
oscillating closed universe that is either non-singular to the past or the
future (or both) also has infinite action, as does a closed universe that
undergoes a single bounce at finite radius. These examples illustrate how
the avoidance of a curvature singularity generally leads to a singularity in
the action.

The imposition of compact topologies on flat or open universes, for example
the flat and open Friedmann universes or the Bianchi type universes, \cite%
{Ash, BK, barr4}, does not produce models with finite action because
although the space integral is finite in eq. (\ref{s}), these models will
expand for ever and create a divergence in the time integration to the
future. For an explicit example of the compact, $T^{3}$, Bianchi type I
model displaying the expected indefinite future expansion which approaches
isotropy in the presence of a perfect fluid, see \cite{top}.

\subsection{Fluid cosmologies}

For homogeneous and isotropic closed universes, we need to consider the
behaviour of the matter action contribution in eq. (\ref{s}). For perfect
fluid models with equation of state $p=(\gamma -1)\rho $ in a Friedmann
universe with scale factor $a(t)=t^{2/3\gamma }$ near the initial
singularity (the final singularity behaviour near $t_{f}$ is just a linear
time translation of this, via $t\rightarrow t_{f}-t$), where $t$ is the
comoving proper time, we have

\begin{eqnarray}
\tint R\sqrt{-g}d^{4}x &\propto &(\gamma -4/3)t^{(2-\gamma )/\gamma },\gamma
\neq 0,2,  \label{2} \\
\tint L_{m}\sqrt{-g}d^{4}x &\propto &\frac{1}{2}\tint (3p-\rho )\sqrt{-g}%
d^{4}x\propto t^{(2-\gamma )/\gamma },\gamma \neq 0,2.  \label{3}
\end{eqnarray}

Importantly, we note that for $\gamma =2$ the $t^{(2-\gamma )/\gamma }$
factor is replaced by $\ln (t)$ which diverges as $t\rightarrow 0$. The $%
\gamma =0$ case is de Sitter or anti-d Sitter spacetime and the action also
diverges because the range of $t$ integration is infinite, as in non-closed
universes. When $\gamma >2$ the action also diverges as $t\rightarrow 0$ and
so finite universal action also excludes the so called ekpyrotic models \cite%
{turok} which behave like \ $\gamma >2$ perfect fluid cosmologies as $%
t\rightarrow 0$ discussed in ref. \cite{coley}. Also excluded are
oscillating cosmologies \cite{dab}, and loop quantum gravity cosmologies
that experience a bounce and past eternal inflationary scenarios like those
in ref. \cite{borde}, whose past geodesic incompleteness was well known from
the properties of the steady state universe \cite{HE, BT2}.

\subsection{Scalar-field cosmologies}

If we have a scalar field, $\phi $, with self-interaction potential $V(\phi
)\geq 0$, in a closed Friedmann universe then we see that if the kinetic
part of the scalar field action, $L_{m}\propto (d\phi dt)^{2}$ dominates on
approach to a singularity (or $V=0$) then we will have $a(t)\propto t^{1/3}$
and $\phi \propto \ln (t)$ and the action diverges as $\tint t^{-2}\times
t^{\ }dt$ $\propto \ln (t)$ as $t\rightarrow 0$ just like in the $\gamma =2$
fluid model. In the borderline case where the kinetic and potential energy
densities are proportional, $\dot{\phi}^{2}\propto V\propto \exp [-\lambda
\phi ]$, we have the asymptotic solution (which is the $k=0$ exact solution) 
\cite{barr1}:

\begin{equation}
\phi \propto \frac{2}{\lambda }\ln (t),\text{ \ \ }a(t)\propto t^{2/\lambda
^{2}}.  \label{exp}
\end{equation}%
Hence the matter action term is proportional to $\tint \dot{\phi}%
^{2}a^{3}dt\propto \tint t^{-2+6/\lambda ^{2}}dt\propto t^{6/\lambda ^{2}-1}$
and this diverges as $t\rightarrow 0$ only when $\lambda ^{2}>6.\ $Scalar
field cosmologies that create a bounce as $t\rightarrow 0$, as discussed
refs \cite{star, barr2, page, schmidt} for $V\propto \phi ^{2}$, are also
ruled out by the finite action principle because they produce a $%
t\rightarrow \infty $ divergence in $S$.

\subsection{Anisotropic and inhomogeneous cosmologies}

If \ we extend our consideration to homogeneous and anisotropic universes
then similar results hold. The Bianchi IX models and their axisymmetric Taub
counterparts with fluids and $S^{3}$ spatial topology have $\sqrt{-g}\propto
a^{3}\propto t$ for the geometric-mean scale factor\ and $\rho \propto
a^{-3\gamma }\propto t^{-\gamma },$ on approach to singularities, therefore

\begin{eqnarray}
S_{m} &\propto &(\gamma -4/3)\tint \rho tdt\propto t^{2-\gamma },\text{ \ }%
\gamma \neq 2,  \label{an1} \\
S_{m} &\propto &\ln (t),\gamma =2.  \label{an2}
\end{eqnarray}%
However, if the cosmology is dominated by anisotropy on approach to either
singularity then the gravitational action will diverge because, with a
Kasner-like vacuum-dominated asymptote, $a(t)\propto t^{1/3}$, and

\begin{equation}
S_{g}\propto \tint R\sqrt{-g}d^{4}x\propto \tint t^{-2}tdt\propto \ln (t),
\label{an3}
\end{equation}

which diverges on approach to the singularity, the anisotropy mimicking the
behaviour of a $\gamma =2$ 'anisotropy' fluid, \cite{mis, barr3}. This will
be the case for the most general Bianchi type IX universes: $S_{g}$ will
diverge as $t\rightarrow 0$ for all $\gamma <2$ fluids and $S_{m}$ and $%
S_{g} $ will diverge when $\gamma =2$. The $\gamma >2$ cases reduce to the
isotropic universe situation studied above and also all have divergent
actions.

The other permitted compact topology for a closed universe which permits a
maximal hypersurface (ie an expansion maximum) \cite{BT2, BGT, barr7} is $%
S^{2}\times S^{1},$ which characterizes the Kantowski-Sachs universes \cite%
{KS, Komp}. The asymptotes are again Kasner-like and the same results hold
for fluid and scalar field sources as for the $S^{3}$ topologies.
Inhomogeneity does not alter these results and the $S^{3}$ Tolman-Bondi
models \cite{kras} and the $S^{3}$ or $S^{2}\times S^{1}$ Szekeres dust
models \cite{szek, kras} all have finite action, whereas the closed $p=\rho $
inhomogeneous closed universes with massless scalar field found by Belinskii 
\cite{bel, kras} have infinite actions like other $\gamma =2$ fluid
cosmologies (although these solitonic solutions are slightly peculiar
because the scalar field depends only on time).

In all anisotropic and inhomogeneous cosmologies containing black body
radiation, magnetic and electric fields, and Yang-Mills fields, we find that
the matter action, $S_{m}$, is finite. The latter two types of source
require anisotropy to be present. These are acceptable finite action matter
sources in the universe, as we would expect.

\ \ If inhomogeneities are added then a density inhomogeneity spectrum of
the power-law form $\delta \rho /\rho \varpropto M^{-q}$ with mass scale $M$
produces divergent metric fluctuations $\delta g/g\varpropto M^{2/3-q\text{ }%
}$of a non-Friedmann type on small scales, as $M\rightarrow 0$ for $q>2/3.$%
This would lead to large anisotropies and divergent action and so is
excluded. On large scales we cannot be so conclusive because we require a
closed universe and the inhomogeneity spectrum will not extend to infinite
mass values. The initial singularity for $q\leq 2/3$ will be quasi-isotropic
with no divergence of the action because of dominant anisotropies.

\subsection{Sudden singularities}

Finite-time singularities of the 'sudden' sort were first introduced into
relativistic cosmology in ref.\cite{bgt} in order to sharpen the conditions
needed to ensure the recollapse of closed universes with $S^{3}$ or $%
S^{2}\times S^{1\text{ \ }}$topologies, and then defined and explored in
detail by the author in \ refs. \cite{sudd1,sudd2,sudd3,sudd4,sudd5,sudd6}.
Other finite-time singularities where the Hubble parameter evolves as $%
H(t)=h_{1}(t)+h_{2}(t)(t-t_{s})^{\lambda \text{ }}$were subsequently defined
and investigated, see refs. \cite{odin, singh} for a classification. The
finite-time singularities occurring are $\lambda <-1$ for 'big rip', $%
-1<\lambda <0$ for 'sudden' (also known as type III), $0<\lambda <1$ for
'type II', and $\lambda >1$ for 'type IV'.

Sudden singularities create no geodesic incompleteness: they are soft
singularities \cite{tip1, kr, soft, soft2} and have been shown to be part of
the general 9-function solution of the Einstein equations in the absence of
an equation of state \cite{soft2}. So, in an ever-expanding universe, we
might integrate the action to future infinity to obtain a singularity.
However, suppose we assume that the past singularity displayed no action
singularity because it was like a perfect fluid Friedmann model and the
evolution did not proceed beyond the finite-time singularity at $t_{s}$.
Then, in the simplest example, at a sudden singularity the quantities $%
a(t),H(t)$ and $\rho (t)$ are finite as $t\rightarrow t_{s}$ but there are
infinities in the pressure, $p$, and the acceleration, $\ddot{a}$, \cite%
{sudd1}. On approach to the sudden singularity as $t\rightarrow t_{s}$ we
have

\begin{equation}
a(t)=\left( \frac{t}{t_{s}}\right) ^{q}(a_{s}-1)+1-(1-\frac{t}{t_{s}}%
)^{n}\rightarrow a_{s}+q(1-a_{s})(1-\frac{t}{t_{s}}),  \label{suddlate}
\end{equation}%
with $1<n<2$, where at early times we can have a standard Friedmann fluid
evolution with

\begin{equation}
a(t)\approx \left( \frac{t}{t_{s}}\right) ^{q},  \label{suddearly}
\end{equation}%
as $t\rightarrow 0$, with $q=1/2$ for radiation domination. Both $a(t_{s})$
and $\dot{a}(t_{s}$) are finite but $\ddot{a}(t_{s})\rightarrow -\infty $
when $1<n<2$, as

\begin{equation}
\ddot{a}\rightarrow q(q-1)Bt^{q-2}-\frac{n(n-1)}{t_{s}^{2}(1-\frac{t}{t_{s}}%
)^{2-n}}\rightarrow -\infty  \label{4}
\end{equation}

Thus assuming finiteness of the space integration in eq. \ref{s} and the
behaviour as $t\rightarrow 0$, we have on approach to the sudden singularity,

\begin{equation}
S_{g}\varpropto \dint\limits_{0}^{t_{s}}\frac{\ddot{a}}{a}\sqrt{-g}%
dt\varpropto -\dint\limits_{0}^{t_{s}}\frac{n(n-1)a_{s}^{2}}{t_{s}^{2}(1-%
\frac{t}{t_{s}})^{2-n}}dt.  \label{sudact1}
\end{equation}%
Hence, we have

\begin{equation}
S_{g}\varpropto \frac{na_{s}^{2}}{t_{s}}(1-\frac{t}{t_{s}})^{n-1},
\label{sudact2}
\end{equation}

and this converges to a finite value as $t\rightarrow t_{s}$ in the sudden
singularity regime with $1<n<2.$ The action will be infinite for finite-time
singularities in the $n<1$ domain.

\subsection{Dark energy and $\Lambda $}

If we return to the scalar field models with exponential potential in a
closed Friedmann universe together with a perfect fluid with equation of
state parameter $2/3<\gamma <2$ then when $\lambda ^{2}>2$ all solutions
start and end with a curvature singularity and have finite action. When $%
\lambda ^{2}<2$ solutions can either recollapse to a singularity or expand
forever, approaching the exact power-law inflationary solutions given above, 
\cite{coley2}. In order to have a description of dark energy that is
consistent with the finite action requirement, we see that we cannot have an
explicit $\Lambda $ term because this will lead to indefinite future
expansion towards de Sitter spacetime \cite{wald, barr5, bouch, starob} and
a divergence of the time integral for the action in eq. (\ref{s}).
Therefore, for finite action, we require the dark energy to be an evolving
scalar field, either explicitly in general relativity as in the action (\ref%
{s}), or via an effective scalar in a modified theory of gravity. This will
allow the closed cosmology eventually to cease to be dominated by the scalar
field and collapse to a future singularity, yielding finite total action. An
unusual example of this sort is the late scenario with the Albrecht-Skordis
potential \cite{alb}\ that was investigated by Barrow, Bean and Magueijo 
\cite{barr6}. A potential of the form

\begin{equation}
V(\phi )=e^{-\mu \phi }P(\phi ),  \label{as}
\end{equation}

where $P(\phi )$ is a polynomial in the scalar field $\phi $ with several
minima, (for example $P(\phi )=A+(\phi -B)^{n}$ in the simplest case), with $%
\mu $ constant, will allow the scalar field to get caught in a succession of
local minima as it rolls down the steep exponential 'cliff'. If the field
comes to rest in a local minimum then $\dot{\phi}=0$ there, and the
expansion dynamics will inflate \cite{alb}. The effect of the crenellations
created by the polynomial $P(\phi )$ is a sequence of accelerating expansion
episodes that can end in non-accelerating expansion or collapse to a future
curvature singularity. However, the $\phi $ field can overshoot or tunnel
through the barrier at a minimum leading to different versions of this
inflation \cite{barr6}. In a closed universe it is possible for this
sequence of inflations to end in collapse to a future singularity and the
universe then has finite total action. A variety of other scenarios are
possible for closed universes to accelerate transiently before recollapsing
to a future curvature singularity. We require the dark energy to display
this evolutionary behaviour culminating in a future curvature singularity
and the universe to be closed.

\section{\protect\bigskip Modified gravity}

When higher-order terms, for example of quadratic and higher orders in the
scalar curvature, \cite{BO}, or higher-order matter terms \cite{board} on
the right-hand side of the field equations, are added to the
Einstein-Hilbert action we expect it to be easier to create an action
singularity at an initial or final singularity of a closed universe. The
finite action principle is at its most powerful when confronting action
singularities in higher-order lagrangian theories. One class of examples
derives from the choice of gravitational lagrangian \cite{BO},

\begin{equation}
L_{g}=f(R),  \label{f}
\end{equation}

whose variation along with the matter action gives the field equations that
generalise Einstein's, :

\begin{equation}
f_{R}R_{ab}-\frac{1}{2}fg_{ab}+f_{R}^{;cd}(g_{ab}g_{cd}-g_{ac}g_{bd})=T_{ab},
\label{f1}
\end{equation}

where $f_{R}=df/dR.$

\subsection{Power-law lagrangians: f(R)=R$^{1+\protect\delta }$}

Barrow and Clifton have examined the case of power-law lagrangians, that
reduce to general relativity as the constant $\delta \rightarrow 0,$ in some
detail \cite{bclif, clif}. It is possible to find exact isotropic and
anisotropic solutions \cite{bclif2, bclif3} of the field equations in vacuum
and with perfect fluids. For equation of state $p=(\gamma -1)\rho $, the
zero-curvature Friedmann metric has the exact solution for the scale factor

\begin{equation}
a(t)=t^{2(1+\delta )/3\gamma },\text{ \ }\gamma \neq 0,  \label{power}
\end{equation}

\bigskip where the constraint on $\delta $ and the value of the scalar
curvature are given by

\begin{eqnarray}
(1-2\delta )\left[ 2-3\delta \gamma -2\delta ^{2}(1+3\gamma )\right] &=&%
\frac{1}{4}(1-\delta )\gamma ^{2}\rho _{c},  \label{power1} \\
R &=&3\delta (1+\delta )t^{-2},  \label{power2}
\end{eqnarray}

where $\rho _{c}$is the Friedmann critical density\footnote{%
When $\gamma =0$, the solution becomes the exact de Sitter metric with $%
a-\exp (nt)$ with the constraint
\par
$3(1-2\delta )n^{2}=(1-\delta )\rho _{c}.$}. This zero-curvature solution is
the behaviour of the closed models as they approach their initial and final
singularities \cite{bclif, clif}. As in general relativity ($\delta =0$),
there are simple deductions from the finite action principle. The universe
must have compact space sections and have finite total lifetime, from
initial to final singularity. Integrating out the space part of the action
in (\ref{s}), we have for the $\gamma \neq 0$ behaviour on approach to a
singularity (the $\gamma =0$ solution has infinite action and is excluded
here):

\begin{equation}
S_{g}\propto \tint R^{1+\delta }\sqrt{-g}dt\propto \tint t^{-2(1+\delta
)}t^{2(1+\delta )/3\gamma }dt\propto t\ ^{[2\delta (1-3\gamma )+2-3\gamma
]/3\gamma },  \label{power3}
\end{equation}

which diverges as $t\rightarrow 0$ if $\gamma >0$ and

\begin{equation}
2\delta (1-3\gamma )+2-3\gamma <0,  \label{del}
\end{equation}%
Hence, in the radiation case ($\gamma =4/3$) we exclude $\delta >-1/3$ and
for dust ($\gamma =1$) we exclude $\delta >-1/4.$

In the anisotropic case of the generalised Kasner vacuum solution in this
theory found in refs. \cite{bclif2, bclif3}, $\sqrt{-g}\propto t^{1+2\delta
} $ and so

\begin{equation}
S_{g}\propto \tint R^{1+\delta }\sqrt{-g}dt\propto \tint t^{-2(1+\delta
)}t^{(1+2\delta )\ }dt\propto \tint t^{-1}dt\propto \ln (t),  \label{5}
\end{equation}%
and there is a logarithmic singularity in the action as $t\rightarrow 0$
just as in general relativity.

Another two exact Friedmann exact solutions of this theory for
zero-curvature Friedmann universes are

\begin{eqnarray}
a(t) &=&t^{\frac{\delta (1+2\delta )}{1-\delta }},  \label{6} \\
a(t) &=&t^{1/2},  \label{7}
\end{eqnarray}%
independent of the matter content $\gamma $. Note that the second solution
does not require radiation to be present, as was first found in refs. \ \cite%
{bher1, bher2} and exists in vacuum, as do the radiation solutions with
non-zero curvature $a\left( t)\varpropto (t-kt^{2}\right) ^{1/2}$.

\subsection{Quadratic gravity}

If we choose a lagrangian of quadratic form, so

\begin{equation}
L_{g}=R+AR^{2},\text{ \ \ }A\text{ constant,}  \label{quad}
\end{equation}%
then if the dynamics on approach to a singularity are close to Friedmann in
the case of a closed universe containing perfect fluid then the contribution
of the $R^{2}$ lagrangian term to the universal action will be

\begin{equation}
S\propto \tint R^{2}\sqrt{-g}dt\propto \tint t^{-4}t^{2/\gamma }\ dt\propto
t^{2/\gamma -3}.  \label{quadS}
\end{equation}%
This diverges on approach to the initial or final singularity for $\gamma
>2/3,$which includes the physically interesting cases of dust, radiation,and
stiff fluid and the divergence in the latter case is stronger that in the
general relativity case, as expected from the $R^{2}$ term. Likewise any $%
R^{n}$ addition to the Einstein-Hilbert lagrangian will create a stronger
divergence in the action on approach to a singularity.

\subsection{Gauss-Bonnet lagrangian}

Consider the general quadratic lagrangian without the $\Lambda $ term:

\begin{equation}
L_{g}=R+\alpha R^{2}+\beta R_{ab}R^{ab}+\mu R_{abcd}R^{abcd}  \label{GB}
\end{equation}%
where $R_{ab}$ is the Ricci tensor and $R_{abcd}$ is the Riemann tensor, and 
$\alpha ,\beta $ and $\mu $ are arbitrary constants. If there is an
isotropic and homogeneous cosmological model with scale factor $a=t^{n}$ on
approach to an initial singularity at $t=0$, then the total gravitational
action is \cite{BT1988}

\begin{equation}
S_{g}=\frac{2n(1-2n)t^{3n-1}}{3n-1}+\frac{4n^{2}t^{3(n-1)}}{n-1}\left\{
3\alpha +\beta +\mu +n(n-1)(12\alpha +3\beta +2\mu )\right\} .  \label{gb}
\end{equation}

We recognise the first term on the right-hand side as the general relativity
contribution from the variation of $R$, which diverges as $t\rightarrow 0$
for $n\leq 1/2$, as shown above. When $\alpha :\beta :\mu =1:1:-4$ the
quadratic terms in \ref{gb} create a complete divergence that leads to no
contributions to the field equations in four spacetime dimensions, so the
field equations are the same as for general relativity. However, the
quadratic terms still contribute to the gravitational action and can create
a divergence as $L_{quad}\varpropto n^{3}t^{3(n-1)}$. Therefore there is an
action singularity as $t\rightarrow 0$ when $n\leq 1$, and hence for all
perfect fluid models with $\gamma \geq 2/3$ (since $n=2/3\gamma $ for
perfect fluid Friedmann solutions). The special Gauss-Bonnet combination $%
\alpha :\beta :\mu =1:1:-4$ is therefore excluded in Friedmann models even
though it does not affect the field equations. We expect anisotropic models
to be excluded also. The finite action principle only allows (\ref{gb}) to
be finite on approach to a singularity for the special (radiation-like) case:

\begin{equation}
n=1/2,\beta =-2\mu /5  \label{8}
\end{equation}%
The higher-order terms in the Lovelock lagrangian \cite{love} in more than
four spacetime dimensions could also be examined for action singularities.
We expect them to be singular in an analogous way since they include higher
powers of the curvature invariants.

\subsection{Brans-Dicke}

Brans-Dicke (BD) cosmological models of Friedmann type fall into two classes
depending on the boundary condition imposed on the BD scalar field \cite{gur}%
. As discussed in detail in ref. \cite{BT1988}, the 'Machian' models which
are matter-dominated near the initial and final singularity in closed
universes have finite action except when $\gamma =2$. By contrast, in the
general case when the cosmologies are dominated by the BD scalar field on
approach to the singularities, there are infinite actions. This is confirmed
by analysis of the general vacuum solutions which are approached at the
singularity by the scalar field (vacuum)-dominated solutions: all have
infinite action and behave as if they are $\gamma =2$ general relativistic
cosmologies, displaying a logarithmic singularity $S_{g}\varpropto \ln (t)$
as $t\rightarrow 0.$ Horndeski lagrangians \cite{horn} can also have their
cosmological conclusions tested again the finite action requirements and
again will be challenged by the presence of higher-order curvature and
scalar-field terms and possibly singularities at finite times for particular
choices of coupling that are linked to the well-posedness of the initial
value problem \cite{thor, pap}.

\section{Conclusions}

We have reconsidered and extended the cosmological predictions that can be
made under the assumption that the total action of the universe is finite.
This is an interesting cosmological constraint because attempts to avoid
cosmological singularities in curvature invariants appear to lead generally
to singularities in the gravitational action. Specifically, we have shown
that the simple ansatz that the total action of the universe be finite has a
large number of powerful consequences. We require the following properties
to be possessed by a universe with finite action:

a. There must be compact spatial sections (ie a closed universe) but compact
spatial sections in flat and open universes created by topological
identifications lead to infinite actions and are excluded.

b. There must be Initial and final singularities (ie a finite total lifetime
for the universe).

c. The universe cannot be dominated by massless scalar fields or the kinetic
energy of self-interacting scalar fields or a $p=\rho $ perfect fluid on
approach to an initial or final singularity

d. The universe can contain perfect fluids with $-1<p/\rho <2$ on approach
to initial and final singularities

e. The universe cannot display a bounce' or indefinite cyclic behaviour to
the past or the future.

f. An ekpyrotic scenario with an effective fluid obeying $p/\rho >2$ in a
closed, flat or open universe is ruled out.

g. Any loop quantum gravity model experiencing a bounce and indefinite past
or future evolution is ruled out.

h. The Einstein static and steady state universes are ruled out along with
past eternal inflating universes and future ever-expanding eternally
inflating universes.

i. Anisotropies (notably those of Kasner or Mixmaster type) cannot dominate
the dynamics of the universe on approach to initial and final singularities.
Inhomogeneities cannot dominate if they induce dominant anisotropies of
Kasner or Mixmaster type. For example, this excludes density inhomogeneity
spectra versus mass scale with $\delta \rho /\rho \varpropto M^{-q}$, for $%
q>2/3.$

j. Higher-order lagrangian theories of gravity are significantly constrained
because the action diverges faster than in general relativity when powers of 
$R$ exceeding unity dominate on approach to a singularity. For example,
quadratic lagrangians are excluded with fluids satisfying $p/\rho >-1/3$.
Gravitational lagrangians with $L_{g}=R^{1+\delta }$ and $p=(\gamma -1)\rho $
fluids have infinite actions on approach to a singularity when $2\delta
(1-3\gamma )+2-3\gamma <0.$We also find that the Gauss-Bonnet quadratic\
lagrangian combination causes an action singularity even though it does not
contribute terms to the field equations.

k. Dark energy cannot be provided by a simple cosmological constant, which
would create an action singularity to the future. The universe cannot be
asymptotically de Sitter as $t\rightarrow \infty $. The dark energy
therefore needs to be an evolving energy density (or effective energy
density) in a closed universe that produces collapse to a future singularity
after a de Sitter-like accelerated expansion phase of finite duration. An
Albrecht-Skordis potential for a scalar field in a closed universe that
recollapses to the future is an admissible example with finite action \cite%
{alb, barr6}. There are many others.

l. Scalar-tensor theories like Brans-Dicke and its generalisations cannot
have cosmological solutions that are dominated by the scalar-field on
approach to singularities. They must track the special matter-dominated
('Machian') solutions \cite{BD, gur, J, will} and must have $p<\rho .$

In conclusion, we have shown that the requirement that the total
gravitational and matter actions of the universe be finite produces a number
of powerful predictions about the geometrical and topological structure of
the universe, its early expansion dynamics, the equation of state of its
material content, the presence of scalar fields, the nature of the dark
energy, and the allowed form of modifications to general relativity. We have
confirmed the constraints found in \cite{BT} with some more recent
applications added in points(a), (b), and (e). We have established new
consequences of finite action in points (c), (d), (f), (g), (h), (i), (j)
and (k), and indicated further theoretical developments in modified gravity
that will reveal new conclusions in the context of Lovelock and Horndeski
actions.

\bigskip \emph{Acknowledgements }

I would like to thank Frank Tipler for earlier collaboration and thank J.
Magueijo, J-L. Lehners and G. Gibbons for helpful discussions and
communications. Support by the Science and Technology Facilities Council of
the UK (STFC) is acknowledged.

\end{document}